\documentstyle[aps,floats,epsf,twocolumn]{revtex}
\begin{document} \draft

\title{Generalized London free energy for high-$T_c$
vortex lattices}

\author{Ian Affleck$^{1,2}$, Marcel Franz$^{3,*}$
and M.H. Sharifzadeh Amin$^1$}
\address{$^1$Department of Physics and Astronomy
and $^2$Canadian Institute for Advanced
Research, University of British Columbia,\\ Vancouver, BC, V6T
1Z1, Canada\\
$^3$Department of Physics and Astronomy, McMaster University,
Hamilton, ON, L8S 4M1, Canada
\\ {\rm(\today)}} 
%
\address{~
\parbox{14cm}{\rm
\medskip
We generalize the London free energy to include four-fold
anisotropies which could arise from d-wave pairing or
other sources in a tetragonal material. We use this  simple model
to study vortex lattice structure and discuss neutron scattering,
STM, Bitter decoration and $\mu$SR experiments.
}}
\maketitle
\narrowtext
The London free energy provides a very simple way of studying the
vortex lattice in an extreme type II superconductor. The conventional
isotropic model\cite{Tinkham}
predicts a hexagonal vortex lattice with an arbitrary orientation
relative to the ionic lattice. Recent neutron scattering\cite{Keimer} and
STM\cite{Maggio} experiments on high-$T_c$ compound YBa$_2$Cu$_3$O$_{7-\delta}$
(YBCO) revealed vortex
lattices with centered rectangular symmetry and a specific orientation
with respect to the ionic lattice. It has been proposed that
this effect can arise from additional quartic derivative terms
in the Ginzburg-Landau (G-L) free energy
\cite{Hohenberg,Takanaka,Ichioka2,Won}
or, alternatively, from
including two or more order parameters (such as $d$ and $s$)
in the G-L free energy with
derivative mixing terms reflecting the ionic lattice symmetry
\cite{Volovik,Berlinsky,Franz,Ichioka}.  Such models predict interesting
effects in the
behavior of the various order parameters in the vortex lattice.
However, these models contain a large number of unknown parameters and
are rather cumbersome to work with numerically.  Another approach
\cite{Yip,Wang}
to the macroscopic effects  of $d$-wave pairing takes into account the
generation of quasiparticles near the gap nodes due to current flow and
thermal excitation.  This leads to a non-linear relationship between
supercurrent and superfluid velocity which becomes singular at $T\to 0$.
 
The purpose of this
letter is to present a simple and general approach to these effects based on a
generalization of the London free energy to include anisotropy of four-fold
symmetry, characteristic of a tetragonal ionic lattice.  The number of new
parameters is far smaller than in the G-L approach (a reasonable model
contains only one new parameter) and numerical simulations are considerably
easier. It provides a useful model to study vortex lattice structure, pinning
by twin boundaries and the magnetic field distribution measured in
$\mu$SR experiments. The model is suitable to study the intermediate
field region $H_{c1}\ll H\ll H_{c2}$ which is experimentally most relevant but
traditionally difficult to handle within the G-L theory.  Furthermore, this
approach can be extended to $T=0$ where G-L theory breaks down
and the supercurrent becomes singular.

We now present a derivation of the generalized London model,
starting from a G-L free energy density with both $d$ and $s$ order
parameters\cite{Volovik,Joynt,Ren}:
\begin{eqnarray}
f= \alpha_s|s|^2 &+& \alpha_d|d|^2 +
   \gamma_s|{\vec\Pi} s|^2 + \gamma_d|{\vec\Pi} d|^2  + f_4 + h^2/8\pi
 \nonumber \\
   &+&\gamma_v\bigl[ (\Pi_y s)^*(\Pi_y d) - (\Pi_x s)^*(\Pi_x d) + {\rm c.c.}
\bigr].
\label{fgl}
\end{eqnarray}
Here $\vec \Pi \equiv -i \nabla -e^*\vec A/\hbar c$ and $f_4$ contains the
quartic terms. We shall consider a case of a $d$-wave superconductor in
which $s$ identically vanishes in zero magnetic field. In finite field
$(H>H_{c1})$ a small $s$-component  with a highly anisotropic spatial
distribution is nucleated in the vicinity of a vortex giving rise to
non-triangular equilibrium lattice structures\cite{Berlinsky,Franz}.
Our strategy will be to simplify free energy (\ref{fgl}) by integrating
out this $s$-component in favor of higher order derivative terms in $d$.
In this process some short length-scale information on the order parameter
is lost but  the magnetic field distribution is described accurately.
Using its Euler-Lagrange equation $s$ can be expressed to the leading
order in $(1-T/T_c)$ as
\begin{equation}
s = (\gamma_v/\alpha_s)(\Pi_x^2-\Pi_y^2)d.
\end{equation}
Substituting this into $f$ gives the leading derivative terms in $d$
of the form:
\begin{equation}
f = \gamma_d[|\vec \Pi
d|^2-(\gamma_v^2/\gamma_d\alpha_s)|(\Pi_x^2-\Pi_y^2)d|^2] + \dots
\label{f4}
\end{equation}
Various additional corrections to the
free energy are obtained from integrating out $s$ more accurately,
taking into account the $\gamma_s|\vec \Pi s|^2$ term and  quartic
terms.  However these all involve higher powers of $\vec \Pi$ or other
terms that will not concern us. The coefficient of the second term has
dimensions of (length)$^2$; we will write it in the form $\epsilon
\xi^2/3$ where $\epsilon\equiv 3(\alpha_d\gamma_v^2/\alpha_s\gamma_d^2)$
is a dimensionless parameter which controls
the strength of the $s$-$d$ coupling and
$\xi\equiv\sqrt{\gamma_d/|\alpha_d|}$ is the G-L coherence length.
We henceforth assume
$\epsilon \ll 1$.  As we remark below, neutron scattering and STM
experiments probably support this assumption.
We note that a term of the form  $|(\Pi_x^2-\Pi_y^2)d|^2$
 could arise without invoking $s$-$d$ mixing from a systematic
derivation of higher order terms in the G-L free energy starting with a
BCS-like model and taking into account the square symmetry of the Fermi
surface \cite{Hohenberg,Takanaka,Ichioka2,Feder}.

The free energy of Eq. (\ref{f4}) is not bounded below,
exhibiting runaway behavior for rapidly varying $d$-fields.  This is in
fact cured by keeping additional higher derivative terms that also
arise from integrating out $s$.  Stability occurs for
$\gamma_v^2<\gamma_s\gamma_d$.  In fact, the approximation of Eq.
(\ref{f4}) will be sufficient for our purposes, yielding a local
minimum which we expect would become a global minimum upon including
the additional terms.

We now assume that the penetration depth $\lambda \gg\xi$.  We may then
assume that $|d(\vec r)|\approx d_0$, the zero field equilibrium
value, almost everywhere in the vortex lattice, except within a
distance of O($\xi$) of the cores.  This gives the London free energy,
\begin{eqnarray}
f_L = (1/8\pi)(\vec B)^2&+&\gamma_dd_0^2\{\vec v^2-(\epsilon
\xi^2/3)[(v_x^2-v_y^2)^2 \nonumber \\
&+&(\partial_yv_y-\partial_xv_x)^2]\},\label{LFE}
\end{eqnarray}
written in terms of the superfluid velocity,
\begin{equation}
\vec v \equiv \nabla \theta -(e^*/\hbar c)\vec A,
\end{equation}
where $\theta$ is the phase of $d$.

The corresponding London
equation, obtained by varying $f_L$ with respect to $\vec A$, is:
\begin{eqnarray}
{c\over4\pi}\nabla \times \vec B &=& \left({2e^*\over\hbar c}\right)
\gamma_dd_0^2\{\vec v -{\tiny{2\over 3}}\epsilon \xi^2[(\hat y v_y-\hat x
v_x)(v_y^2-v_x^2) \nonumber \\
&-&(\hat y \partial_y-\hat x
\partial_x)(\partial_yv_y-\partial_xv_x)]\}.\label{j}
\end{eqnarray}
For many purposes it is very convenient  to
express $\vec v$ in terms of $\vec B$ and its derivatives, and then
substitute this expression for $\vec v$ back into $f_L$, giving an
explicit expression for $f_L$ as a functional of $\vec B$ only.  For
$\epsilon = 0$ this gives
\begin{equation}
\vec v^{(0)}=\nabla \times \vec B /B_0,
\label{v0}
\end{equation}
where $B_0\equiv \phi_0/2\pi \lambda^2$  is of order $H_{c1}$ ($\phi_0\equiv 2\pi \hbar c/e^*$ is the flux quantum) and
\begin{equation}
f_L^0 = (1/8\pi )[\vec B^2+\lambda_0^2(\nabla \times \vec
B)^2].
\end{equation}
Here the penetration depth, for $\epsilon = 0$ is
$\lambda_0^{-2}=8\pi \gamma_d(e^*d_0/\hbar c)^2$.  It is presumably not
possible to solve Eq. (\ref{j}) in closed form for $\vec v$
as a function of $\vec B$ for $\epsilon\ne 0$.
However, this can be done readily in a
perturbative expansion in $\epsilon$.  The first order correction is:
\begin{eqnarray}
\vec v^{(1)}&=&
(2\epsilon \xi^2/3)\{(\hat y v_y^{(0)}-\hat x v_x^{(0)})[(v_y^{(0)})^2-
(v_x^{(0)})^2] \nonumber \\
&-&(\hat y \partial_y-\hat x
\partial_x)(\partial_yv_y^{(0)}-\partial_xv_x^{(0)})\}
\end{eqnarray}
with $\vec v^{(0)}$ given by Eq. (\ref{v0}).  The
London free energy density, up to O($\epsilon $) is then:
\begin{equation}
f_L = f_L^0+
{\epsilon \lambda_0^2\xi^2\over 8\pi}[4(\partial_x\partial_yB)^2+
((\partial_xB)^2-(\partial_yB)^2)^2/B_0^2].
\label{f}
\end{equation}
Note that we could have arrived at a similar conclusion by simply
writing down all terms allowed by symmetry in $f_L$, expanding in
number of derivatives and powers of $B$.  Square anisotropy is first
possible in the fourth derivative terms.  In principle, we should
 also include all isotropic terms to order $B^4$  and $\nabla
^4$.  However, assuming that these have small coefficients, they will
not be important.  This result can also be obtained from considering
generation of quasi-particles near gap nodes in a $d$-wave
superconductor \cite{Yip}, in a range of temperature and field where
the supercurrent can be Taylor expanded in the superfluid
velocity\cite{Affleck}.  More generally, the quadratic and quartic
terms in (\ref{f})  have independent coefficients.

The corresponding London equation is obtained by varying $f_L$ with
respect to $\vec B(\vec r)$. For $B$ along the $z$ direction one obtains
\begin{equation}
[1-\lambda_0^2\nabla^2+4\epsilon
\lambda_0^2\xi^2(\partial_x\partial_y)^2]B -\epsilon Q[B]=0,
\label{london}
\end{equation}
where
\begin{eqnarray}
Q[B]&=&2\lambda_0^2\xi^2 B_0^{-2}
[(\partial_x^2-\partial_y^2) B+\partial_xB\partial_x-\partial_y
B\partial_y] \nonumber \\
&\times &[(\partial_x B)^2-(\partial_y B)^2]
\end{eqnarray}
is the non-linear term arising from the last term in Eq. (\ref{f}).
\begin{figure}
\epsfxsize=8.5cm
\epsffile{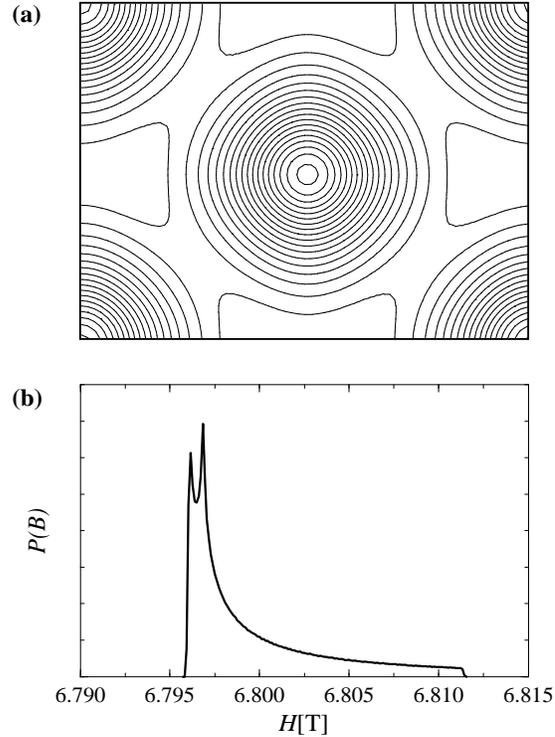}
\caption[]{
a) Distribution of magnetic field in a vortex lattice for $\epsilon=0.3$
and $H=6.8$T, leading to an angle of $\beta\simeq74^\circ$. We use
$\lambda_0=1400$\AA\  and $\kappa\equiv\lambda_0/\xi=68$. b) Corresponding
$\mu$SR lineshape.}
\label{fig:1}
\end{figure}

To get a feeling for the effect of the extra terms, consider a weak
field which depends only on $x$ or else only on $(x+y)$. The solution of
the linearized London equation [(\ref{london}) without the last term]
gives an exponentially decaying field with $\lambda = \lambda_0$ for
variation along the $x$-axis but:
\begin{equation}
\lambda = \lambda_0\left[{1\over 2}+ \sqrt{{1\over
4}-{\epsilon \xi^2\over \lambda_0^2}}\right]^{1/2},
\end{equation}
for variation at 45$^\circ$ to the crystal axis\cite{remark1}.
The penetration depth is longer along the crystal axis.

\begin{figure}
\epsfxsize=8.5cm
\epsffile{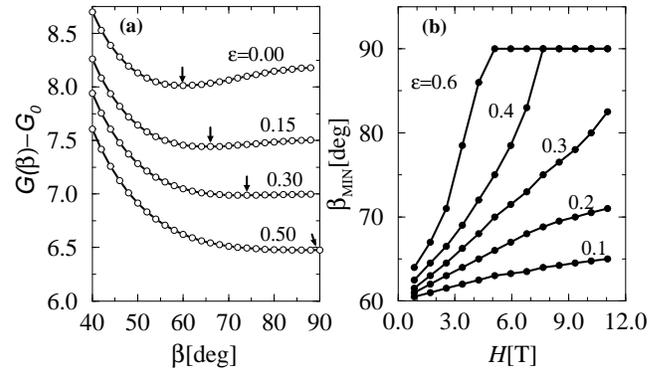}
\caption[]{
a) Gibbs free energy as a function of $\beta$ for the same parameters
as Fig.\ \ref{fig:1} and  various values of $\epsilon$. Arrows indicate
positions $\beta_{\rm MIN}$ of the minima and $G_0\equiv -H^2/8\pi$.
b) Equilibrium angle $\beta_{\rm MIN}$ as a function of $H$ for several
values of $\epsilon$.}
\label{fig:2}
\end{figure}

To determine vortex lattice structure we
insert source terms  $\sum_j\rho(\vec r-\vec r_j)$ at the vortex core
positions, $\vec r_j$, on the right hand side of Eq. (\ref{london}).
The source terms reflect the topological winding of the phase angle
and the reduction  of the order parameter in the core\cite{Tinkham}.
A commonly used phenomenological form is\cite{Brandt}:
\begin{equation}
\rho (\vec r) = (\phi_0/2\pi\xi^2 )e^{-r^2\xi^2/2}.
\end{equation}
It is straightforward to solve these equations numerically
for the vortex lattice
by an iterative method.  We find that the quartic term makes a
negligible contribution.  (Contrary to naive expectation, it doesn't
become more important with increasing applied field because the field
becomes nearly constant in the vortex lattice when the applied field is
large.) Thus to an excellent approximation one may neglect $Q[B]$ in the
London equation (\ref{london}) and the magnetic field may be written
explicitly as:
\begin{equation}
B(\vec r) = \bar B\sum_{\vec k}{e^{i\vec k \cdot \vec r}e^{-k^2\xi^2/2}\over
1+\lambda_0^2k^2+4\epsilon \lambda_0^2\xi^2(k_xk_y)^2}.
\end{equation}
Here the sum is over all wave-vectors in the reciprocal lattice and $\bar
B$ is the average field.  The lattice constant is determined by the
condition that $\bar B\Omega=\phi_0$ where $\Omega$ is the area of the
unit cell.  The lattice symmetry is determined by minimizing the Gibbs
free energy $g_L=f_L-H\bar B/4\pi$.  We
find that the flux lattice has centered rectangular symmetry, with
principal axes aligned with the ionic crystal lattice, with an angle
$\beta$ between unit vectors which depends on $\epsilon$ and the
magnetic field. An example of such a centered rectangular lattice is shown
in Fig.\ \ref{fig:1}(a).
In agreement with earlier results within G-L\cite{Franz}
and Eilenberger\cite{Ichioka} formalisms individual vortices are elongated
along the crystalline axes.  Figure \ref{fig:2}(a) shows the dependence of
Gibbs
free energy on $\beta$ for various values of $\epsilon$ at fixed applied
field $H=400B_0\simeq 6.8$T. For $\epsilon =0$ minimum occurs for
$\beta_{\rm MIN} = 60^\circ$,  corresponding to a hexagonal lattice. As
$\epsilon$ increases $\beta_{\rm MIN}$ continuously increases and
for sufficiently large $\epsilon$, the flux lattice
becomes tetragonal with $\beta_{\rm MIN} = 90^\circ$.
For $\beta_{\rm MIN} \neq 90^0$,
there are always two solutions, related by a $90^\circ$ rotation,
in which the
long axis of the centered rectangle is aligned with either the $x$ or $y$
axis.  The degeneracy is much larger for $\epsilon = 0$, when the flux
lattice may have an arbitrary orientation relative to the ionic
crystal lattice.

Dependence of
$\beta_{\rm MIN}$ on the applied field for various values of  $\epsilon$ is
displayed in Fig.\ \ref{fig:2}(b). Clearly the anisotropic term becomes more
important at larger fields. Our perturbative elimination of $\vec v$ in
favor of $B$ breaks down when $\epsilon$ and $H$ are sufficiently large that
$\beta_{\rm MIN}$ differs significantly from $60^\circ$.  Furthermore, we
might expect higher order corrections to (\ref{LFE}) to be important in
this regime.   By fitting   Fig. \ref{fig:2}(b)
to experimental data on tetragonal materials such as
Tl$_2$Ba$_2$CuO$_{6+d}$ (once such data
become available) one can directly assess the magnitude of $\epsilon$, the
only unknown parameter in the model.

Our analysis can be easily extended to take into account effective
mass (i.e. penetration depth) anisotropy.  In a simple one-component
G-L model, the derivative term is generalized to:
\begin{equation} f = \sum_{i=x,y,z}\gamma_i|\Pi_i d|^2.\end{equation}
We restrict our attention to fields along the $z$-axis.  Then the
anisotropy can be removed by a rescaling of the $x$-coordinate and a
corresponding rescaling of the magnetic field.  The coherence length
and penetration depth anisotropies are the same:
$\xi_y/\xi_x=\lambda_x/\lambda_y$.  We will make the simplifying
assumption that the higher derivative and mixed derivative terms in
$f$ are also simply modified by a rescaling by a common factor.  It
then follows that the flux lattice shape is obtained by stretching
along the $x$-axis by the factor $\lambda_x/\lambda_y$.   We now obtain two
possible vortex lattices, both of centered rectangular symmetry, aligned with
the ionic lattice, with different angles, $\beta$. (Relaxing our simplifying
assumption may split the degeneracy between these two lattices.)
On the other hand, when $\epsilon = 0$, we may rotate the hexagonal
lattice by an arbitrary angle before stretching. This gives
an infinite set of oblique lattices with arbitrary orientation.

To compare theory with YBCO we should take into account twin
boundaries which may also tend to align the vortex lattice by pinning
vortices to the twin boundaries, at $\pm 45^0$ to the $x$-axis.  This
effect competes with alignment to the ionic lattice which we have been
discussing.  Only in the special case of a square vortex lattice does a
line of vortices occur at $\pm 45^0$.  If this is not the case, and if
pinning by twin boundaries is significant, then we should expect that
the vortex lattice will align with the ionic lattice far from twin
boundaries but will be deformed in the vicinity of a twin boundary in
an effort to align itself with the twin boundary. On the other hand,
for $\epsilon =0$, the vortex lattice would remain aligned with the twin
boundaries everywhere except within vortex lattice domain boundaries
which necessarily exist roughly midway between the twin boundaries.

Neutron scattering experiments on YBCO \cite{Keimer}
suggest that the vortex lattice is well-aligned with the twin boundaries and
is close to being centered rectangular (the ratio of lattice constants is
about 1.04) with $\beta \approx 73^0$, with weak dependence on $H$. This
corresponds to a rotation away from alignment with the ionic lattice by
$9^\circ$.  Four different orientational domains, related by reflection in the
$(1,1,0)$  axis and $90^0$ rotation were reported.
STM imaging of the YBCO vortex lattice also suggests that the
(highly disordered) lattice has approximately centered rectangular symmetry
with $\beta \approx 77^\circ$. However, no evidence for the $9^\circ$ tilt
into alignment with the twin boundaries was reported.

These neutron scattering results can be  rather well fitted\cite{Walker} by
the basic London model ($\epsilon =0$) with mass anisotropy.  There is a
unique stretched hexagonal lattice which is aligned with either the
$(1,1,0)$ or $(1,\bar 1,0)$ twin boundaries.
For $\lambda_x/\lambda_y=1.5$, a value roughly consistent with
infrared and microwave experiments, this lattice has about the right
shape. Taking into account the two crystallographic domains (related by
interchanging $\lambda_x$ and $\lambda_y$) there are all together four
vortex lattice domains, as seen experimentally.  The experimental fact
that the vortex lattice appears to be well aligned with the twin
boundaries suggests that the tendency to align with the ionic lattice
 is small.   No evidence for a  bending of the vortex lattice (by
$9^0$) into alignment with the ionic lattice far from the twin boundaries
has so far been found.

Low field Bitter decoration data on YBCO \cite{Dolan} show vortex
lattice geometry with a very small distortion from hexagonal, consistent with
a much smaller anisotropy $\lambda_x/\lambda_y=1.11-1.15$. One may be tempted
to attribute this apparent field dependence of $\beta$ to the effects
discussed above in connection with Fig.\ \ref{fig:2}(b).
An alternative, and perhaps more likely explanation, is a poor
quality of samples used in the Bitter decoration experiments that may have
resulted in partial washing out of the $a$-$b$ anisotropy otherwise present
in clean crystals\cite{Timusk}.

$\mu SR$ experiments measure the field distribution
$P(B)=(1/\Omega )\int \delta[B-B(\vec r)] dx dy$.  This is shown in
Fig.\ \ref{fig:1}(b).  For $\beta \neq 60^\circ$, $B(\vec r)$ has
two inequivalent saddle points leading to two peaks in $P(B)$.
$P(B)$ is unaffected by effective mass anisotropy, as can be shown by the
rescaling transformation, mentioned above.  Existing $\mu SR$ experiments show
only a single peak \cite{Sonier}.

The weak field dependence of $\beta$, the alignment with twin boundaries and
the single peak in $P(B)$ all suggest that $\epsilon$ is small in YBCO and
that the normal London model, together with twin boundary pinning, provides
a good fit to the data.  Bitter decoration data do not fit well into
this overall picture and further experimental work, preferably on untwinned
YBCO or other tetragonal superconductors, will probably be
necessary to clarify the importance of square lattice anisotropy in
high-$T_c$ superconductors.

The general approach to vortex lattices introduced here may be extended
to low temperatures, but then the free energy takes a quite different
form which is non-analytic in $B$ at $T\to 0$ \cite{Yip,Affleck}.

We would like to thank A. J. Berlinsky, R. E. Kiefl and J. E. Sonier for 
helpful discussions.  This research was supported by NSERC and the CIAR.



\begin{references}
\bibitem{remark1} For $\epsilon\neq 0$ the secular equation for $\lambda$
has two distinct roots, the larger of which is identified as a physical
penetration depth.
\end{references}
\end{document}